\documentclass[showpacs,pre,amsmath,twocolumn,amssymb,floatfix]{revtex4}

\usepackage{epsfig}
\usepackage{graphicx,psfrag}
\usepackage{dcolumn}
\usepackage{bm}

\newcommand     {\beq}[1]         { \begin{equation} #1 \end{equation}
}
\newcommand     {\beqa}[1]        { \begin{eqnarray} #1 \end{eqnarray}
}

\newcommand     {\EP}             { \varepsilon }
\newcommand     {\SI}             { \sigma }

\begin{document}
\title{Damage process of a fiber bundle with a strain gradient} 
\author{Ferenc Kun\footnote{Electronic
address:feri@dtp.atomki.hu} and S\'andor Nagy}
\affiliation{
Department of Theoretical Physics,
University of Debrecen, P.\ O.\ Box:5, H-4010 Debrecen, Hungary}
\date{\today}

\begin{abstract}
We study the damage process of fiber bundles in a wedge-shape
geometry which ensures a constant strain gradient. To obtain the wedge
geometry we consider the three-point bending of a bar, which is
modelled as two rigid blocks glued together by a thin elastic
interface. The interface is discretized by parallel fibers with
random failure thresholds, which get elongated when the bar is bent. 
Analyzing the progressive damage of the system we show that the strain
gradient results in a rich spectrum of novel behavior of fiber
bundles. We find that for weak disorder an
interface crack is formed as a continuous region of failed fibers. 
Ahead the crack a process zone develops which proved to shrink with
increasing deformation making 
the crack tip sharper as the crack advances. For strong disorder,
failure of the system occurs as a spatially random sequence of
breakings. Damage of the fiber bundle proceeds in bursts whose size
distribution 
shows a power law behavior with a crossover from an exponent 2.5 to 2.0
as the disorder is weakened. The size of the largest burst increases as a power
law of the strength of disorder with an exponent 2/3 and saturates for
strongly disordered bundles.
\end{abstract}
\maketitle

\section{Introduction}

The damage and fracture of disordered materials is an important scientific and
technological problem which has attracted an intensive research during
the past years \cite{hh_smfdm,chakrab_beng_book_1997,zapperi_alava_statmodfrac}. 
Theoretical and experimental studies have revealed that at the 
beginning of the loading process of highly disordered materials, first
micro-cracks nucleate randomly, 
covering the entire volume of the specimen without any spatial
correlations \cite{hh_smfdm,chakrab_beng_book_1997,zapperi_alava_statmodfrac,extension_fbm_lnp_kun}. Approaching the critical load, localization occurs
resulting in a single growing crack along which
the specimen falls apart. In a composite
system of two solid blocks glued together along an interface, the
damage usually concentrates along the weak plane of the glue \cite{frank_ferenc_2005,frank_plastic,Knudsen_breaksurf_pre_2005,delaplace_ijss_1999}. 
Loading such composites,
interface crack propagation occurs which has also been found to be a
complex sequence of crack growth and arrest with interesting spatial
and temporal fluctuations
\cite{delaplace_ijss_1999,batrouni_intfail_pre_2002,frank_ferenc_2005,Knudsen_breaksurf_pre_2005,roux_damint_physicaa_1999,maloy_waitingtime_prl,alava_peeling_epl2006}.
The crackling noise accompanying the
failure of disordered systems (bulk or interface cracking) can be
recorded in the form of a complicated trail of  
signals whose analysis provides important information about the
microscopic dynamics of damaging
\cite{hansen_distburst_local_1994,kloster_pre_1997,hansen_crossover_prl,garcimartin_prl79_1997,guarino_exp_acoust_epjb_1998,maloy_waitingtime_prl,salminen_paper_acoust_prl_2002}.

Fiber bundle models (FBM) are one of the most important theoretical
approaches to the progressive damage of disordered materials. During
the last decade FBMs have provided a deep insight into the collective
nature of the microscopic
dynamics and statistical properties of degradation phenomena. 
Recently, FBMs have also been applied to study interfacial failure of glued
solid blocks under shear loading
\cite{frank_plastic,frank_ferenc_2005} and wear
\cite{Knudsen_breaksurf_pre_2005,delaplace_jem_2001,delaplace_ijss_1999}. 
Interesting novel results have been obtained on the temporal and spatial
fluctuations of local breakings which precede macroscopic failure, and
on the analogy of fracture with phase transitions and critical
phenomena \cite{frank_plastic,frank_ferenc_2005,Knudsen_breaksurf_pre_2005,delaplace_jem_2001,delaplace_ijss_1999}. 
In this paper we
study the damage process of a fiber bundle in a wedge-shape geometry,
which provides a constant strain gradient of fibers. To obtain a
simple representation of the geometry and loading conditions, we
consider a bar subject to three-point bending. 
The bar is modelled as two rigid blocks coupled together by an elastic
interface which is then discretized by a bundle of parallel
fibers. Deformation and damage of the bar is concentrated in the 
interface resulting in a linear deformation profile of fibers, while
the two blocks remain intact. Besides interfacial
failure, the model provides the mean field limit of the
failure of disordered materials under three-point bending.  
Varying the amount of disorder of fibers, we can control
the strength of non-linearity before macroscopic failure, and hence,
the type of fracture (brittle-quasi-brittle) of the bundle
\cite{sornette_prl_78_2140,hansen_crossover_prl,frank_cutoff}. 
We focus on the progressive damage of the fiber bundle analyzing the damage
profile, crack formation, and bursts of local breakings.  
We find that for weak disorder an interface crack is formed by a
continuous region of failed fibers.  Ahead the interface crack, a
process zone develops which proved to shrink with increasing
deformation making the crack tip sharper. For strong disorder the
failure of the bundle occurs due to a spatially random sequence of local
breakings. Very interestingly we find that the size distribution of
bursts is a power law whose 
exponent shows a crossover from exponent 2.5 to 2.0 when the strength
of disorder is lowered. 

We demonstrate that the novel results of our model calculations
are the consequence of the strain gradient; in the homogeneous case of
zero gradient our model recovers all recent results of FBMs with
varying threshold disorder
\cite{sornette_prl_78_2140,hansen_crossover_prl,frank_cutoff,kloster_pre_1997,hansen_lower_cutoff_2005,hansen_lower_cutoff_2006}.
Our results imply that interfacial 
fracture problems can lead to novel universality classes of
breakdown phenomena.

\section{The model}
\begin{figure}
\psfrag{xx}{{\Large $\delta$}}
\psfrag{ll1}{\Large$l_1$}
\psfrag{lli}{\Large$l_i$}
\psfrag{aa}{{\Large $b$}}
\psfrag{bb}{{\Large $a$}}
\psfrag{ll0}{{\Large $l_0$}}
\psfrag{FF}{{\Large $F$}}
\epsfig{bbllx=0,bblly=0,bburx=335,bbury=185,file=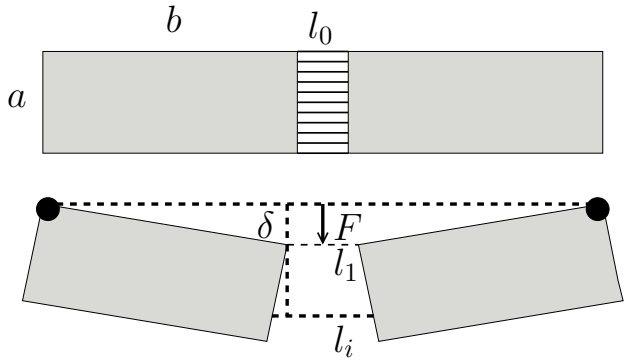,
  width=8.0cm}
\caption{The geometrical layout of the model. The two rigid blocks of
side length $a$ and $b$ are glued together by an interface of width
$l_0$ which is discretized in terms of elastic fibers. The specimen
suffers deflection $\delta$ under the action of the external force $F$
exerted in the middle of the bar. The wedge shaped opening of the
interface results in a linear deformation profile of fibers. 
\label{fig:layout}}
\end{figure}

In order to obtain a fiber bundle with a linear deformation profile,
we construct a simple model for the loading of an elastic bar of 
rectangular shape by
an external force exerted perpendicular to the longer side of the bar
in the middle. For simplicity, in the model the bar is composed of two
rigid blocks of side lengths $a$ and $b$ which are glued together by
an elastic interface of width $l_0$, where $l_0 \ll b$ holds, see Fig.\
\ref{fig:layout}. The interface region can deform and suffer
breaking under deflection of the specimen while the two rigid blocks
remain intact. Bending of the specimen is performed such that the two 
blocks undergo rigid rotation about their outer upper corner
concentrating the deformation in the interface layer. We discretise
the interface in terms of elastic fibers of number $N$ and
length $l_0$ which are placed equidistantly between the two blocks.
The fibers do not have bending rigidity, they can undergo only
stretching deformation characterized by the same value of the Young
modulus $E$. During the bending of the specimen, the fibers can
support only a finite deformation, {\it i.e.} if 
the local deformation $\EP_i$ of fiber $i$ exceeds a threshold value
$\EP_i^c$ the fiber breaks and a 
micro-crack nucleates in the interface. The disordered properties of
the material are represented by the randomness of the breaking
thresholds $\EP_i^c$, which are independent identically
distributed random variables with a probability density $p(\EP)$
and cumulative distribution $P(\EP^c)=\int_0^{\EP^c}p(x)\mathrm{d}x$.
The rigidity of the two rotating blocks implies that
the macroscopic deformation of the specimen can be characterized by a
single variable $\delta$ which denotes the deflection of the middle of the
bar from the original position, see Fig.\ \ref{fig:layout}.

It can be seen in Fig.\ \ref{fig:layout} that under bending of the
specimen the interface opens resulting in an increasing elongation of
fibers from top to bottom. The actual length of fibers $l_i$ can be
expressed as a function of $\delta$ as
\beq{
l_i = l_1+2\delta\frac{a}{b}\frac{i-1}{N-1}, \ \ \ \ \ \ i=1,\ldots ,N
\label{eq:li}
}
where $l_1=l_0+2\left(b-\sqrt{b^2-\delta^2}\right)$ is the length of fiber
index $i=1$ at the top of the bar. It follows that also the 
elongation $\Delta l_i$ and longitudinal strain $\EP_i$ of fibers increase
linearly as a function of their position $i$ 
\beqa{\label{eq:delta}
\Delta l_i &=& 2b-2\sqrt{b^2-\delta^2}+2\delta\frac{a}{b}\frac{i-1}{N-1}, \\
\EP_i      &=& \frac{\Delta l_i}{l_0}.
\label{eq:delta_li}
}
Equilibrium of the system is obtained when the total momentum of
forces with respect to the clamping points is zero.
During the deformation process, those fibers which exceed their
threshold value break, i.e. they are removed from the interface. 
Since $1-P(\EP_i(\delta))$ is the probability that the interface element of
index $i$ remained intact under the externally imposed deformation
$\delta$, based on the equilibrium condition, the constitutive
equation $\sigma(\delta)$ of the deflected bar can be cast in the form
\beqa{
\label{eq:fgen}
\sigma(\delta)&=&\frac1{NL}\sum_{i=1}^N\left[\delta +\sqrt{b^2-\delta^2}\frac{a}{b}
\frac{(i-1)}{(N-1)} \right] \\ \nonumber
&\times & \left[1-P(\EP_i(\delta))\right]E\EP_i(\delta), 
}
where $L=2b+l_0$ is the overall length of the bar and the sum goes
over all the fibers. On the right hand side $\EP_i(\delta)$ should be
substituted from Eqs.\ (\ref{eq:delta},\ref{eq:delta_li}).
The above equations describe the macroscopic response of a fiber
bundle which has a linear deformation profile. In the following, for
the explicit calculations the geometrical parameters were set as
$a=1$, $b=2.5$, and $l_0=0.1$.

\begin{figure}
\epsfig{bbllx=30,bblly=15,bburx=485,bbury=415,file=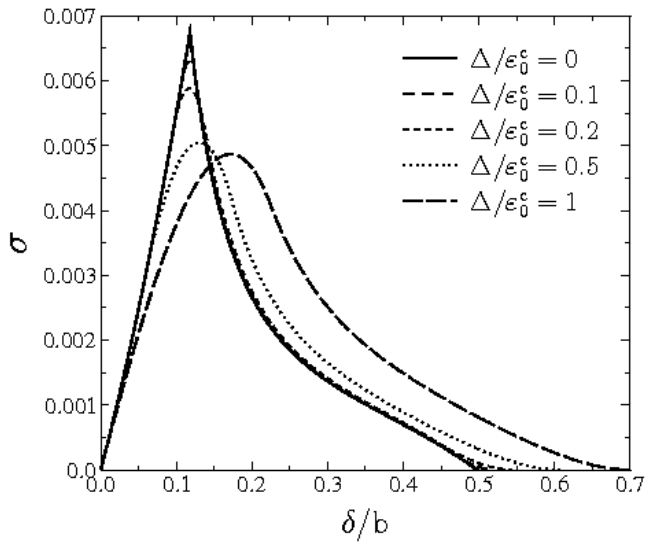,
  width=8.3cm}
\caption{ The  constitutive curve $\sigma(\delta)$ of the interface
composed of $N=10000$ fibers varying the width of the threshold
distribution $\Delta$ at the fixed value of $\varepsilon_0^c=0.01$.
For $\Delta =0$ the constitutive curve $\sigma(\delta)$ has a 
sharp peak  which gets rounded and develops into a
quadratic maximum when $\Delta$ is increased.
\label{fig:dist}}
\end{figure}

The amount of disorder of the failure thresholds $\EP_i^c$ has a substantial effect
on the macroscopic response of the fiber bundle $\sigma(\delta)$. In the limiting
case of zero disorder, {\it i.e.} when all the fibers have the same
breaking threshold $\EP_i^c=\EP_0^c$, the failure of the bundle starts
at the bottom of the interface where the stretching deformation is the highest and proceeds
upwards as $\delta$ is increased. It can be seen in Fig \ref{fig:dist} that
the corresponding constitutive curve is sharply peaked. The critical deformation
$\delta_c$ defined by the peak position corresponds to the instant
of the first fiber breaking $\EP_0^c=\EP_N(\delta_c)$. Beyond the peak
stress, $\sigma$ rapidly decreases due to the gradual
breaking of fibers as $\delta$ increases.
In order to study how the behavior of the system changes
when the amount of disorder of fibers is varied, we
consider a uniform distribution for the breaking thresholds over an
interval of width $2\Delta$ centered at the value $\EP_0^c$
\beq{
P(\EP^c) = \frac{\EP^c-(\EP_0^c-\Delta)}{2\Delta}, \ \ \ \ \ \ \EP_0^c-\Delta <
\EP^c < \EP_0^c+\Delta. 
\label{eq:disorder}
}
The strength of disorder of the breaking thresholds is characterized
by $\Delta$, while $\EP_0^c$ sets the scale of fiber strength. The
width $\Delta$ can be varied over the interval $[0,\EP_0^c]$ where the
limits 
$\Delta = 0$ and $\Delta = \EP_0^c$ corresponds to zero disorder and the
strongest disorder, respectively.
Fig.\ \ref{fig:dist} shows that increasing $\Delta$, the peak of the
constitutive curve gets more 
and more rounded and develops into a quadratic maximum. The maximum
of $\sigma(\delta)$ is preceded by a longer and longer non-linear
regime due to the breaking of fibers, so that for $\Delta \rightarrow
\EP_0^c$ the linear behavior prevails only for small deformations $\delta
\rightarrow 0$. On the microlevel this process is accompanied by the
randomization of the breaking sequence of fibers along the interface,
{\it i.e.} for $\Delta \neq 0$ fibers do not simply break in the
decreasing order of their index $i$ (from bottom to top of the bar). 

\section{Simulation techniques}
\label{sec:simul}
The complete constitutive curve of the system presented in Fig.\
\ref{fig:dist} can only be recovered by deformation controlled
loading. When $\delta$ is controlled externally, the local load on the
fibers is solely determined by the externally imposed deformation,
so that there is no load redistribution after fiber breaking. 
Under stress controlled conditions, the breaking of fibers is followed
by the redistribution of load over the intact ones. 
Due to the wedge shape of the deformed interface, at a given
external load $\sigma$, the load on the fibers linearly increases from
top to bottom. It has the consequence that in spite of the rigidity of
the two solid blocks, the load redistribution following fiber failure differs
from the usual equal load sharing (ELS) approximation commonly used
for the study of parallel bundles of fibers
\cite{zapperi_alava_statmodfrac,extension_fbm_lnp_kun,chakrabarti_phasetrans,sornette_prl_78_2140,kun_epjb_17_269_2000}.
The rigid surfaces, 
however, ensure that the load is redistributed globally
in such a way that the excess load received by an intact fiber depends
on its position along the interface but not on its distance from the
failed one. This implies that no stress enhancement arises in the
vicinity of the failed fibers as in the case of 
local load sharing approximation (LLS) of fiber bundles
\cite{zapperi_alava_statmodfrac,raul_varint_2002,frank_cutoff}. Our
fiber bundle model provides the mean field limit of the damage and
fracture of disordered materials under three-point bending
conditions and also represents an interesting interface rupture
problem.  

In order to analyze the microscopic damage mechanism of FBMs with a
constant strain gradient,
we worked out an efficient simulation technique for a sample where
the interface is composed of $N$ fibers with breaking thresholds
$\EP_i^c, i= 1, \ldots , N$ sampled from the probability distribution
Eq.\ (\ref{eq:disorder}). 
Substituting the breaking thresholds $\EP_i^c$ on the left hand side
of Eq.\ (\ref{eq:delta}) and inverting it for $\delta$, we can determine
the value of the macroscopic deformation parameter
$\delta_i^c=\delta(\EP_i^c,i), i= 1, \ldots , N$ at which the fibers break. Of
course, $\delta_i^c$ is 
a function of both the position of the fiber $i$ along the interface and
the local breaking threshold $\EP_i^c$. During the loading process the
fibers break in the increasing order of their critical macroscopic
deformation $\delta_i^c$, which can be a randomized sequence of the fibers'
position $i$. The
computer simulation of the loading process proceeds as follows: after
generating the breaking thresholds of fibers $\EP_i^c$ we determine
the corresponding critical deflections $\delta_i^c$ and sort 
them into increasing order. The constitutive curve of
the sample can be simply obtained by calculating the load needed to
achieve the deformation $\delta_i^c$ after the breaking of the first $i-1$
fiber remaining only $N_{intact} = N-(i-1)$ intact
elements. Between the breaking of the $(i-1)th$ and the $ith$ fibers,
the constitutive equation of  
the system takes the form
\beqa{
&\sigma& =
\frac{E}{l_0L}\left\{2\delta\left[b-\sqrt{b^2-\delta^2}\right]N_{intact}
\right. \label{eq:no_breaking} \\
  &+& \left[\delta^2\frac{4a}{b}-2ab+2a\sqrt{b^2-\delta^2}\right]\frac{1}{N-1} \nonumber
      \sum_{j=1}^{N}{'} (j-1) \\ \nonumber
  &+&
\left. \frac{2\delta a^2}{(N-1)^2b^2}\sqrt{b^2-\delta^2}\sum_{j=1}^{N}{'}(j-1)^2
\right\}, \nonumber
}
where the prime indicates that the summation is restricted to indices
of intact fibers (which are not necessarily consecutive integers). 
Note that in Eq.\ (\ref{eq:no_breaking}) the value of $\delta$ falls in the
range $\delta_{i-1}^c < \delta < \delta_i^c$.

Performing stress controlled experiments, 
after the breaking of a fiber the deformation
of the specimen can freely change resulting in a redistribution of
load over the intact fibers. The excess load taken up by the intact
fibers can give rise to further fiber failures which may trigger an
entire avalanche of breakings. This avalanche either stops and
the bar becomes stable under the externally imposed load, or it spans
the entire interface and the specimen breaks into two pieces (the
entire bundle ruptures). In order
to study numerically this microscopic breaking process, in the
simulations first we
increase the deformation $\delta$ such that a single fiber breaks,
{\it i.e.} $\delta=\delta_1^c$ with index $i_1$.
Then the load needed to maintain this 
deformation $\sigma$ is calculated from Eq.\ (\ref{eq:no_breaking})
for $N_{intact} = N$ fibers. After the breaking of fiber $i_1$ its
load has to be redistributed over the remaining $N-1$ fibers. In order
to determine the load of intact fibers after the removal of the broken
one, we remove fiber $i_1$ on the right hand side of Eq.\
(\ref{eq:no_breaking}) and invert the equation for $\delta(\sigma)$ keeping the
load $\sigma$ fixed. The fibers with threshold values $\delta_{i}^c <
\delta(\sigma)$ break as a consequence of load redistribution. This
iteration has to be repeated under a fixed external load $\sigma$
until the breaking sequence stops or all the fibers break resulting in
a macroscopic failure of the system.

\section{Spatial evolution of damage}
In a bundle of fibers loaded between two parallel rigid plates, due to
the equal load sharing after fiber breaking, the failure of a
fiber is solely determined by its breaking threshold \cite{hansen_distburst_local_1994,kloster_pre_1997,hansen_crossover_prl,raul_burst_contdam}. Hence, fibers
break in a completely random sequence without any spatial
correlations. In our system, however, during the loading process the
fibers break in the increasing order of 
their critical macroscopic deformation $\delta_i^c(\EP_i^c,i)$ which
depends both on the local breaking thresholds $\EP_i^c$ and on the
spatial position $i$ of fibers. 
\begin{figure}
\begin{center}
\epsfig{bbllx=25,bblly=25,bburx=480,bbury=250,file=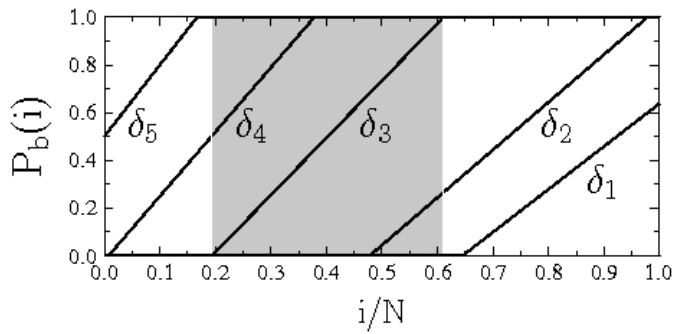,
  width=8.7cm}
 \caption{\small Damage profile, {\it i.e.} the breaking probability $P_b$
of fibers along the interface at five different values of the 
deflection $\delta_1 < \delta_2 <\delta_3<\delta_4<\delta_5$
with $\Delta/\EP_0^c=0.2$ and $b/a = 2.5$. The deflection $\delta_1$
falls in regime $(A)$, $\delta_2, \delta_3, \delta_4$ are in $(B)$,
and $\delta_5$ belongs to $(C)$.
The increase of the slope
of the straight lines indicates the
sharpening of the profile. The shadowed area highlights the process
zone for $\delta_3$.  
}
\label{fig:damprof}
\end{center}
\end{figure}
In the limiting case of zero disorder,
{\it i.e.} $\Delta = 0$ and $\EP_i^c = \EP_0^c$ the critical deformation
$\delta_i^c(\EP_0^c,i)$ is a monotonically decreasing function of $i$,
which implies that the fibers break one-by-one starting from the
bottom. This breaking sequence can be conceived as a crack is
generated which penetrates the
interface upward such that below the crack tip all the fibers are
broken while above it the fibers are intact. Under strain
controlled loading stable crack propagation is obtained gradually
breaking the fibers by the strain increments. Controlling the external
load, however, the onset of crack propagation occurs in an unstable
manner resulting in immediate catastrophic failure when the maximum of
$\SI(\delta)$ is reached (see Fig.\ \ref{fig:dist}).

 Increasing the strength of disorder $\Delta$, the breaking
sequence of fibers determined by $\delta_i^c$ becomes spatially
randomized. At a given deformation $\delta$, the fibers with
$\delta_i^c < \delta$ have already failed. If 
the disorder is not too strong, an interesting
spatial distribution of these broken fibers emerges: starting from the
bottom of the interface a 
continuous region of failed fibers develops forming a
crack. On the opposite side, starting from the top of the interface a
continuous region of 
intact, elongated fibers can be observed. The two regimes are
separated by a process zone, which is a sparse sequence of intact and
broken elements. To illustrate this feature, in Fig.\
\ref{fig:damprof} we show the probability $P_b(i)$ that the fibers are
broken along the interface for several different values of $\delta$ using
the threshold distribution Eq.\ (\ref{eq:disorder}). The
probability $P_b(i)$ that fiber $i$ is broken at the deflection
$\delta$ can be obtained directly from the threshold distribution
$P_b(i) = P(\varepsilon_i(\delta))$. 
The process zone is defined as the regime where for the probability
$P_b$ of fiber breaking $0 < P_b < 1$ holds.
It can be observed in Fig.\ \ref{fig:damprof} that the process zone
sharpens, {\it i.e.} its  
width decreases as the deformation $\delta$ increases which makes the
crack tip sharper as the crack advances. 
\begin{figure}
\begin{center}
\epsfig{bbllx=0,bblly=0,bburx=460,bbury=380,file=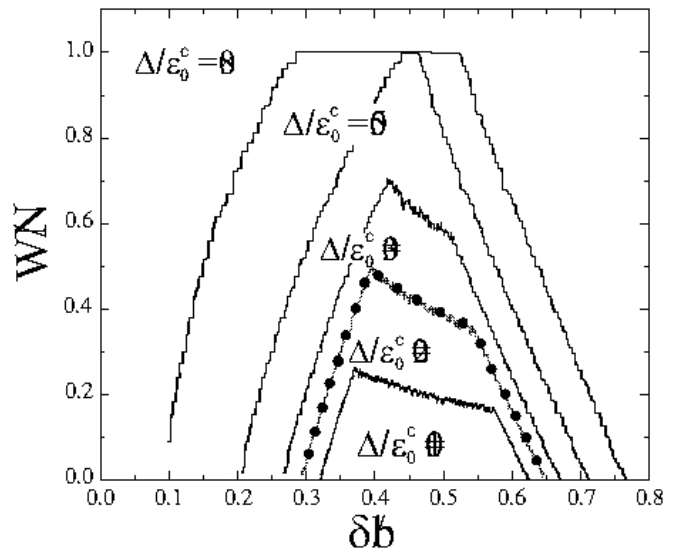,
  width=8.7cm}
 \caption{\small {\it (Color online)} Width of the process zone $W/N$
as a function of the 
deflection $\delta$ of the bar for several values of
$\Delta$. Computer simulations are in a perfect agreement with the
analytic predictions Eqs.\ (\ref{eq:pz1},\ref{eq:pz2},\ref{eq:pz3})
represented by the dots for the specific case of
$\Delta/\EP_0^c = 0.2$. }
\label{fig:proc_zone}
\end{center}
\end{figure}
It can be obtained
analytically that at a given deflection $\delta$ the width $W$ of the
process zone depends both on the strength of disorder $\Delta$  and on
the geometrical extensions $a,b$ of the specimen. For the explicit
calculations, it is worth considering separately the following three
regimes of the damage profile: $(A)$ for the breaking probability at
the bottom of the interface 
it holds that $P_b(i=N) < 1$, {\it i.e.} the crack has not yet
developed ($\delta_1$ in Fig.\ \ref{fig:damprof}); $(B)$ The process
zone is completely contained by the 
interface $P_b(i=1) = 0$ and $P_b(i=N) = 1$ ($\delta_2, \delta_3,
\delta_4$ in Fig.\ \ref{fig:damprof}), and $(C)$ there is no
intact region $P_b(i=1) > 0$ ($\delta_5$ in Fig.\
\ref{fig:damprof}). The width $W$ of the process zone can be obtained
analytically as a function of $\delta$ for the three cases
\beqa{
(A) \hspace*{0.3cm} W &=&
{\displaystyle N\left[1-\frac{\EP_0^c-\Delta-2b+2\sqrt{b^2-\delta^2}}{2\delta}\frac{b}{a}
\right]}, \hspace*{0.3cm} \label{eq:pz1} \\ 
(B) \hspace*{0.3cm} {\displaystyle W} &=&
{\displaystyle\frac{\Delta}{\delta}\frac{b}{a}(N-1), \label{eq:pz2} \hspace*{0.3cm}}
\\  
(C) \hspace*{0.3cm} W &=& {\displaystyle
N\frac{\EP_0^c+\Delta-2b+2\sqrt{b^2-\delta^2}}{2\delta}\frac{b}{a}. \hspace*{0.3cm}}
\label{eq:pz3}
}
We also determined the width of the process zone $W$ numerically for a
system of $10^6$ fibers, which is presented in Fig.\
\ref{fig:proc_zone} together with the corresponding analytic results. 
In Fig.\ \ref{fig:proc_zone} the curves of $W(\delta)$ are composed of
three distinct parts corresponding 
to the regimes $(A)$, $(B)$, and $(C)$ of Eqs.\
(\ref{eq:pz1},\ref{eq:pz2},\ref{eq:pz3}). It 
can be seen that for smaller values of
$\Delta/\varepsilon_0^c$ first $W$ increases and reaches a maximum
where the crack occurs. As the crack advances, the width of the process zone
decreases according to
Eq.\ (\ref{eq:pz2}) and finally, as the tip of the process zone
reaches the top of the interface, $W$ rapidly decreases as given by
Eq.\ (\ref{eq:pz3}). It is interesting to note that for large $\Delta$
values no crack can be identified, {\it i.e.} for the parameter set
used in Fig.\ \ref{fig:proc_zone} the damage profile spans the entire
interface $\mbox{max} [W/N]=1$ when $\Delta/\EP_0^c$ exceeds 0.46.
It follows from the above arguments that the disorder of the
interface can be considered strong if the damage 
proceeds as a spatially random sequence of local breakings without the
formation of a propagating crack. No crack can develop if at the
deformation $\delta$ where the top of the interface may already be
damaged $\EP_1(\delta) > \EP_0^c-\Delta$ the bottom of the interface
may still be intact $\EP_N(\delta) < \EP_0^c+\Delta$.
Making use of Eq.\ (\ref{eq:delta}) and assuming $b \gg
\EP_0^c-\Delta$, the condition of 
strong disorder can be formulated as 
\beq{
\frac{b}{a^2}\Delta^2 + \Delta > \EP_0^c,
}
which implies that the average fiber strength $\EP_0^c$, the width of the
distribution $\Delta$ and the geometrical layout $a, b$ of the specimen
together determine the relevance of disorder. At a given value of $a$,
$b$, and $\EP_0^c$, the crossover point $\Delta^*$ between weak and
strong disorder can be obtained as
\beq{
\Delta^* \approx \frac{a^2}{2b}\left[\sqrt{1+\frac{4b\EP_0^c}{a^2}}-1
\right],
\label{eq:cross}
}
so that for $\Delta > \Delta^*$ no crack is formed, while for $\Delta
< \Delta^*$ crack propagation occurs with a shrinking process zone
ahead the crack tip. For the parameter values of Fig.\ 
\ref{fig:proc_zone} the crossover point $\Delta^* = 0.4633$ was
obtained from Eq.\ (\ref{eq:cross}) in an excellent agreement with the
numerical results.

\section{Bursts of fiber breakings}
In order to characterize the damage process of the fiber bundle under
stress controlled conditions, we determined the distribution of burst
sizes $s$ of fiber breakings 
varying the width $\Delta$ of the disorder distribution. Simulations
were carried out by increasing the external load 
\begin{figure}
\begin{center}
\epsfig{bbllx=10,bblly=30,bburx=585,bbury=280,file=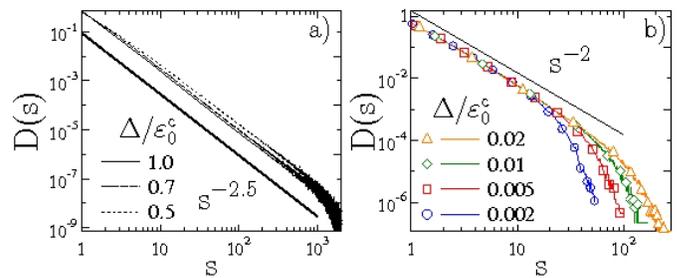,
  width=9.0cm}
 \caption{\small {\it (Color online)} $a)$  Distribution of burst
sizes $D(s)$ varying the 
value of $\Delta$ in a broad range. $b)$ Distributions for the limiting case
of very weak disorder. Simulations were carried out with $N=10^6$
fibers averaging over $10^3$ samples. 
The geometrical layout of the sample was $b/a =2.5$, and the scale
parameter of fibers' strength had the value $\varepsilon_0^c=0.01$.
Straight lines of slope $2.5$ and
$2.0$ are drawn as reference.
}
\label{fig:avalanche}
\end{center}
\end{figure}
to break a single fiber and following the cascading fiber breakings
with the algorithm discussed in Sec.\ \ref{sec:simul}.
The avalanche size distributions $D(s)$ are presented in Fig.\
\ref{fig:avalanche} varying the amount of disorder in a broad range. 
It can be observed in Fig.\ \ref{fig:avalanche} that $D(s)$
has a power law form 
\beq{
D(s) \sim s^{-\tau}
\label{eq:aval_dist}
}
at any finite value of $\Delta$ with an exponential cutoff at large
avalanches. Simulations revealed an interesting 
change of the value of the exponent  $\tau$ as the strength of
disorder is varied: for strong disorder the value of $\tau$ 
coincides with the mean field (equal load sharing) exponent of the
classical parallel bundle of fibers $\tau \approx 2.5$
\cite{hemmer_distburst_jam_1992,kloster_pre_1997}. However, as the
disorder is weakened, the distribution exhibits a crossover to another
power law with a lower exponent $\tau\approx 2.0$. To better
illustrate this effect, in Fig.\ \ref{fig:avalanche}$b)$ burst
distributions are shown separately for the limiting case of very weak
disorder. The numerical results are well described by a power law with
an exponent 2.0.
 
\begin{figure}
\begin{center}
\epsfig{bbllx=35,bblly=445,bburx=365,bbury=735,file=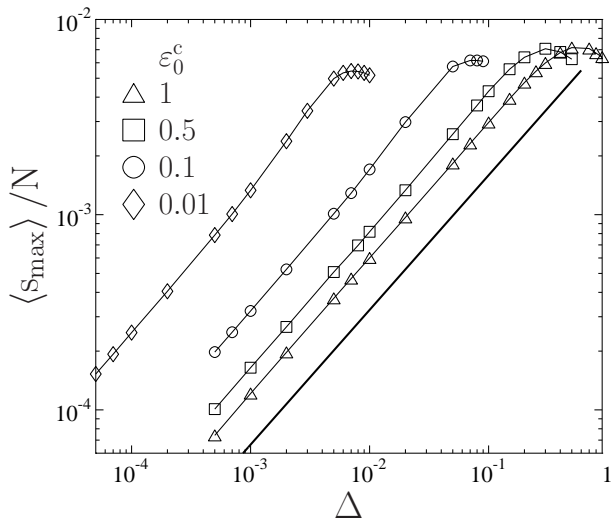,
  width=8.0cm}
 \caption{\small 
Average size of the largest avalanche as a function of $\Delta$ for
several values of $\EP_0^c$. A straight line of slope $2/3$ is drawn
to guide the eye.
}
\label{fig:maxes}
\end{center}
\end{figure}

It is interesting to note that the size of the largest burst $s_{max}$
has a strong dependence on the value of $\Delta$ (see Fig.\
\ref{fig:avalanche}$(b)$), namely, as the strength of disorder is reduced
the largest avalanche decreases. To obtain a quantitative
characterization of 
this effect, Fig.\ \ref{fig:maxes} presents the average size of the
largest bursts $\left< s_{max}\right>$ as a function of $\Delta$,
where a power law dependence is evidenced  
\beq{
\left< s_{max} \right> \sim \Delta^{\alpha}
\label{eq:smax}
}
for the case of weak disorder $\Delta < \Delta^*$. As $\Delta$ exceeds
$\Delta^*$, the largest avalanche $s_{max}$ reaches a maximum and levels
off (see Fig.\ \ref{fig:maxes}). Computer simulations revealed that the exponent $\alpha$ has
a universal value $\alpha = 2/3$ indicated by the 
straight line drawn in Fig.\ \ref{fig:maxes} to guide the eye.
The value of $\Delta$ where $\left<s_{max}\right>$ saturates is in a good
agreement with the corresponding value of $\Delta^*$ estimated from
Eq.\ (\ref{eq:cross}).

It is important to emphasize that in our system no critical disorder
distribution can be identified in the sense defined in Refs.\
\cite{hansen_crossover_prl,hansen_lower_cutoff_2005,hansen_lower_cutoff_2006,frank_cutoff,divakaran_cutoff_2007,divakaran_mixedweibull_2007}.
For equal load sharing fiber bundles the threshold
distribution is considered to be critical if the breaking of the
weakest fiber gives rise to an immediate macroscopic failure of the system. For
uniformly distributed threshold values in the interval $[x_0,1]$ the
distribution becomes critical for $x_0 \to 0.5$
\cite{hansen_crossover_prl,hansen_lower_cutoff_2005,hansen_lower_cutoff_2006,frank_cutoff,divakaran_cutoff_2007,divakaran_mixedweibull_2007}.
It has been pointed out that approaching the critical disorder, the
macroscopic response of the system becomes perfectly brittle and on the
micro-level the size distribution of bursts exhibits a crossover from
a power law of exponent 2.5 to a significantly lower one 1.5
\cite{hansen_crossover_prl,frank_cutoff}.  
Using our terminology, such critical behavior in a bundle of fibers
loaded between two parallel rigid plates, {\it i.e.} in the case of
equal load sharing, should be obtained for 
$\Delta \to \EP_0^c/3$, however, computer simulations revealed a
sudden collapse of our FBM with constant strain gradient
solely at $\Delta =0$. 

\begin{figure}
\begin{center}
\epsfig{bbllx=25,bblly=590,bburx=430,bbury=770,file=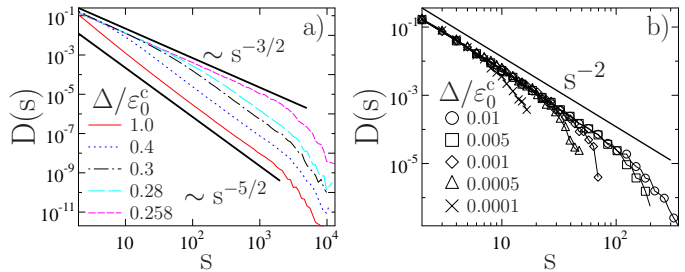,
  width=8.7cm}
 \caption{\small {\it (Color online)} $a)$  Burst size distributions
for the limiting case of zero strain gradient ($a=0.0$). There exists
a finite critical disorder $\Delta_c \approx 0.258$, where a crossover
occurs from the exponent $5/2$ to a lower value $3/2$. $(b)$ At any
finite value of $a>0$ criticality is obtained only for $\Delta \to 0$
and the crossover exponent becomes larger. 
}
\label{fig:homo}
\end{center}
\end{figure}

The reason for the missing critical state in our model is the
inhomogeneity of the load of fibers along the interface. Eq.\
(\ref{eq:delta}) shows 
that at any deflection $\delta$ the load of intact fibers linearly
increases from top to bottom of the interface where the gradient, {\it
i.e.} the strength of inhomogeneity, is determined by the geometry of
the system $a/b$. Setting the cross section of the specimen $a$ to
zero $a=0$, the positional dependence of $\varepsilon_i$ disappears in
Eqs.\ (\ref{eq:delta},\ref{eq:delta_li},\ref{eq:fgen})
and formally all the fibers keep the same load determined by $\delta$. We
carried out computer simulations in the limiting case $a=0$ varying
the strength of disorder $\Delta$. We indeed find
that in this homogeneous case a critical state arises at
$\Delta_c/\varepsilon_0^c \approx 0.258$ such that for $\Delta <
\Delta_c$ a single fiber breaking triggers a catastrophic
avalanche. The corresponding numerical results are presented in Fig.\
\ref{fig:homo}$(a)$, where it can be seen that approaching $\Delta_c$ 
from above, the burst size distribution exhibits a crossover from the
exponent $\tau=5/2$ to a lower value $\tau=3/2$ in agreement with the
predictions of Refs.\
\cite{hansen_crossover_prl,hansen_lower_cutoff_2005,hansen_lower_cutoff_2006}.
The deviation of $\Delta_c$ from $\varepsilon_0^c/3$ arises due to the
non-linear terms of $\delta$ in Eqs.\
(\ref{eq:delta},\ref{eq:fgen}). It can also be observed in the figure
that in the homogeneous case the largest burst $s_{max}$ increases as
criticality is approached in agreement with Ref.\ \cite{frank_cutoff}. 

However, at any finite value of $a$, {\it i.e.} in the presence of a strain
gradient, the picture drastically changes: criticality occurs solely
in the limit $\Delta\to 0$ so that catastrophic failure is always
preceded by avalanches with a power law distribution but the cutoff
avalanche size goes to zero as a power law of $\Delta$ when $\Delta$
is decreased. Fig.\ \ref{fig:homo}$b)$ presents simulation results
obtained with $10^6$ fibers at the value $a=0.01$, which results in a
much lower strain gradient than in Fig.\ \ref{fig:avalanche}. Comparing
the burst size distributions of different values of $\Delta$ for the
homogeneous (Fig.\ \ref{fig:homo}$a)$) and inhomogeneous (Fig.\
\ref{fig:homo}$b)$) cases, a clear difference can be observed. The
higher value of the crossover exponent $\tau = 2$ compared to
$\tau=3/2$ of the homogeneous case shows that in the absence of a
finite critical disorder, the large avalanches are less dominating in
the distributions $D(s)$.

\section{Discussion}
We presented a detailed study of the progressive damage and fracture
of fiber bundles in a wedge-shape geometry which provides a linear
deformation profile for fibers. For a simple representation of the
geometrical and loading conditions of the system, we considered a bar
subjected to three point bending. The bar is composed of two rigid
blocks coupled by an elastic interface which is then discretized in
terms of parallel fibers. We showed that in the limit of zero disorder
of fibers' strength the bundle has a perfectly brittle macroscopic
response, {\it i.e.} under stress 
controlled loading global failure occurs as a sudden collapse of the
system without any precursory activity, furthermore, fibers break in a
completely ordered sequence from bottom to top of the wedge creating an
instable crack with a sharp tip. The relevance of disorder is
determined together by the parameters (mean and width) of the strength
distribution of fibers and by the geometrical layout of
the wedge. We demonstrated that a propagating interface crack can only be
defined for weak disorder. Ahead the crack a process zone is formed
whose width decreases with increasing deformation making the crack tip
sharper as the crack advances. For strong disorder a spatially random
sequence of local breakings occur along the entire bundle. 

The breaking of single fibers can trigger cascades of
breaking events. The size distribution of these bursts is found to be
a power law with an interesting crossover effect as the strength of
disorder is varied: for strong disorder the mean field exponent $\tau
= 5/2$ of equal load sharing fiber bundles is recovered indicating the
complete randomness of the failure process. However, for
weak disorder where a propagating crack with a process zone develops,
a lower exponent $\tau =2.0$ is obtained. In the weak 
disorder regime the largest burst increases as a power law of the
width of the disorder distribution with an exponent $\alpha =
2/3$.  We showed
that in the limit of zero strain gradient our calculations reproduce
the crossover of burst exponents from $5/2$ to $3/2$ predicted
recently. The novel features of our system originate from the finite
strain gradient, {\it i.e.} from the geometrical constraints of fibers
which naturally occur, for instance, at interfacial fracture problems. 
Our model provides also the mean field limit of the damage and
fracture of disordered materials under three-point bending.
The results imply that the statistical physics of interfacial rupture can
reveal novel universality classes of breakdown phenomena. 

Recently, it was found experimentally that the cracking of a bar under
three point bending proceeds in bursts which are characterized by
power law distributions \cite{kun_magnetnoise_prl_2004}. The
experiments showed that the exponents of the amplitude, area and
energy distribution of magnetic emission signals recorded during the
fracture process of ferromagnetic materials are sensitive to the type
of fracture, {\it i.e.} the noise spectra of ductile materials are
characterized by higher exponents than the 
brittle ones. The boundary and loading conditions ensured in the
experiments that the damage localizes to a relatively thin layer of
the specimen giving
rise to a single growing crack so that the crackling noise measured
during the loading process characterizes the crack propagation. 
Note that in our model the crossover to a lower exponent of burst sizes with
decreasing disorder is accompanied on the macro-level by an increasing
degree of brittleness showing that this simple mean field approach can
qualitatively account for the changing properties of crackling noise
observed experimentally \cite{kun_magnetnoise_prl_2004}. 

\begin{acknowledgments}
This work was supported by OTKA T049209 and NKFP-3A/043/04. 
\end{acknowledgments}

\bibliography{../../../FIBER/statphys_fracture.bib}

\end{document}